# Electrostatics and the Assembly of an RNA Virus

Paul van der Schoot[§] and Robijn Bruinsma

Department of Physics and Astronomy

UCLA, Box 951547

Los Angeles, CA 90095-1547

U.S.A.

Abstract:

Electrostatic interactions play a central role in the assembly of single-stranded RNA viruses. Under physiological conditions of salinity and acidity, virus capsid assembly requires the presence of genomic material that is oppositely charged to the core proteins. In this paper we apply basic polymer physics and statistical mechanics methods to the self-assembly of a synthetic virus encapsidating generic polyelectrolyte molecules. We find that (i) the mean concentration of the encapsidated polyelectrolyte material depends on the surface charge density, the radius of the capsid, and the linear charge density of the polymer but neither on the salt concentration or the Kuhn length, (ii) the total charge of the capsid interior is equal but opposite to that of the empty capsid, a form of charge reversal. Unlike natural viruses, synthetic viruses are predicted not to be under an osmotic swelling pressure. The design condition that self-assembly only produces *filled* capsids is shown to coincide with the condition that the capsid surface charge exceeds the *desorption threshold* of polymer surface adsorption. We compare our results with studies on the self-assembly of both synthetic and natural viruses.

PACS: 64.10.+h 82.70.-y 36.20.-r 87.15.Aa

---

[§] On leave from the Faculteit Technische Natuurkunde, Technische Universiteit Eindhoven, The Netherlands.

# I. Introduction

Electrostatics plays a pivotal role in the formation of a virus. A first indication comes from the fact that the reconstitution of an (infectious!) viral particle under *in-vitro* conditions from an aqueous solution containing the viral protein and RNA molecules only succeeds for a certain salinity range and salt composition [1,2]. Structural studies provide more detailed information concerning the role of electrostatics. A virus consists, minimally, of a protein shell (or "capsid") that protects the enclosed RNA or DNA genome. The proteins (or "subunits") that constitute the capsid carry, under physiological *pH* conditions, a typical positive electrical charge of 11 to 13 on a section of the protein that faces the interior (usually, but not exclusively, near the amino terminal [3]). Negative charges tend to be located on the exterior surface of the capsid. The number of proteins per capsid ($N$) equals 60 times the "T-Number", a structural index for viral capsids that adopts certain integer values such as 1, 3, 4, and 7, so the total positive charge of a virus can be substantial. The focus of this paper will be on small (i.e., T = 3 or 4) RNA viruses - like the Polio Virus – which have single-stranded ("ss") genomes with a typical size of the order of 3-4 kilobases (usually partitioned among a few RNA molecules) [4]. At physiological *pH*, the total negative charge of the phosphate groups of the RNA backbone is then about twice the total positive charge of the capsid interior surface, so the virus interior has a net negative charge of the order of (minus) $10^3$. In general, viral particles carry a significant net charge [5], which helps prevent virus-virus aggregation.

The electrostatic repulsion between charged capsid proteins should inhibit viral self-assembly. This is confirmed by studies of the phase diagram of solutions of capsid proteins [6,7]. The fraction of capsid proteins that aggregate into (empty) shells, or other protein clusters, increases with salinity ("salting-out"). For instance, the classical study by Aaron Klug of the phase diagram of tobacco mosaic virus (TMV) subunits found that under physiological conditions only small disk-like oligomers form while for higher salt concentrations cylindrical aggregates appear that actually resemble (empty) TMV capsids [6]. Viral protein aggregation can be analysed in terms of the competition between a salt-dependent electrostatic protein-protein repulsion and a (largely) salt-independent hydrophobic protein-protein attraction [8]. In particular, a recent and quantitative study by Ceres and Zlotnick of the self-assembly of capsid proteins of Hepatitis B Virus (HBV) indicated that, under very carefully controlled conditions, the fraction of aggregated proteins will follow the *law of mass action* of equilibrium thermodynamics [7]. (This seems to be true also for TMV [9].) The formation free energy of the HBV capsids could be obtained this way and was found to be of order 2,000 $k_BT$ per capsid. The measured salt dependence of this formation energy was found to follow the Debye-Hückel theory of screened interactions in aqueous solution, applied to a charged spherical shell [8]. The resulting capsid surface charge density $\sigma$, obtained from a fit to the experimental data, was found to be almost one net electrical charge per nm$^2$, which agrees reasonably well with the nominal "chemical" charge of about 0.4 per nm$^2$ of the capsid proteins [10].

The electrostatic self-repulsion of the genome also has an inhibitory effect on assembly. Light-scattering and small-angle X-ray studies have shown that viral RNA molecules in physiological solution have a significantly lower density than these same molecules inside the viral capsid [11]. In fact, the nucleotide density inside a T = 3 RNA virus is comparable to that of a (hydrated) RNA *crystal* [12]. The thermodynamic work required to compactify the genome against the electrostatic self-repulsion during assembly is not known for the case of the single-stranded ("ss") viral RNA genomes. However, for certain DNA viruses – the bacteriophages – it has been demonstrated that the electrostatic self-repulsion of the genome generates an internal osmotic pressures of the order of tens of atmospheres [13], while the work of compaction is of order $10^3 - 10^4$ $k_BT$ [14], Note that this is comparable to the formation energy of an empty HBV capsid.

Despite all this, electrostatics in general – specifically the electrostatic attraction between capsid and genome – provides the thermodynamic driving force for viral self-assembly, at least for ss RNA viruses [15]. It is important to distinguish at this point *specific* from *non-specific* interactions. Klug showed [6] that the disk-like TMV oligomers that form under physiological conditions actually will assemble into fully infectious TMV viruses once viral RNA molecules are added to the solution. Assembly is initiated by the specific affinity of a certain hairpin-shaped RNA sequence along the viral genome – the so-called "packaging signal" – for the oligomers [16]. After formation of this initial nucleo-protein complex, subsequent assembly of the virus is driven by non-specific affinity between the viral RNA molecule and the capsid proteins. Similar RNA-driven assembly scenarios, where a specific initial RNA-oligomer assembly involving either hairpin packaging signals [12] or tRNA-like structures [17] develops into a fully formed infectious virus, driven by non-specific affinity, are encountered also for spherical viruses, though assembly intermediates are somewhat harder to isolate [18]. The non-specific affinity is usually, though not always, electrostatic attraction between the positive charges of the subunits and the negative charges of the genome [19]. The importance of a purely non-specific, electrostatic thermodynamic driving force for viral assembly was demonstrated early on by the classical studies of Bancroft and collaborators who showed that capsid proteins of certain viruses, such as those of the Cowpea Chlorotic Mottle Virus (CCMV), Brome Mosaic Virus (BMV) and Broad Bean Mottle Virus (BBMV), actually will package *alien* RNA molecules – including purely homopolymeric RNA – and even generic polyelectrolytes [4].

This counterpoint between electrostatic repulsion and electrostatic attraction clearly plays an important role during viral assembly. Apparently, under the right conditions, electrostatic repulsion between subunits is able to prevent assembly of *empty* capsids while electrostatic attraction between subunits and genome molecules is strong enough to overcome this repulsion and allow *filled* and not empty capsids to assemble. The aim of this paper is to apply methods borrowed from polymer physics and the statistical mechanics of self-assembly to examine just what exactly the "right conditions" should be to allow this balancing act. Knowing these conditions should offer a guide for the laboratory synthesis of artificial capsids designed to carry a polymeric cargo. In particular, we would like to establish a theoretical limit on the amount of material that can be encapsidated purely by spontaneous self-assembly as well as the amount of encapsidated material that minimizes the free energy. We will focus here solely on the *equilibrium thermodynamics* of viral assembly. As mentioned, specific interactions dominate formation kinetics whereas non-specific electrostatic interactions dominate the thermodynamic driving force for the growth of the subunit/genome aggregates. The advantage of this focus on the non-specific thermodynamic driving force is that we do not need to concern ourselves with the secondary structure of the RNA molecules and specific packaging signals. To that effect, we will focus on the encapsidation in saline solution of a "toy genome" consisting of flexible, negatively charged, soluble, homo-polymeric polyelectrolytic material by positively charged "protein" units that can aggregate into shells of fixed radius $R$ and surface charge density $\sigma$. The study by Bancroft and co-workers [4] of the encapsidation of homopolymeric RNA by CCMV capsid proteins would provide a specific realization of the model.

The main results of our investigation are as follows.

(i) If the capsid is *permeable* to the polyelectrolyte material, and if the co-assembly takes place under condition of full chemical equilibrium, then the mean internal concentration $\langle\phi\rangle^*$ of encapsidated polyelectrolyte material should equal $6\sigma/\alpha R$ with $\alpha$ the number of charges per Kuhn length, depending neither on the salinity of the solution nor on the Kuhn length of the polyelectrolyte. The concentration profile is characterized by *power-law* behavior. The total charge of the assembly is approximately equal but opposite to the charge of the empty capsid.

(ii) If the capsid is *impermeable* to the polyelectrolyte material, but permeable to water and to salt ions, then a range of packing concentrations $\langle\phi\rangle$ different from $\langle\phi\rangle^*$ is possible. If assembly proceeds sufficiently slowly, then the mean concentration is expected to lie in the interval $1/2\langle\phi\rangle^*$ to $\langle\phi\rangle^*$. The maximum possible concentration of packaged polyelectrolyte material consistent with spontaneous self-assembly is of order $\sigma/\sqrt{Rd}$, with $d \ll R$ the so-called "extrapolation length". The concentration profile is characterized by a uniform "core" and a surface layer of enhanced monomer concentration with a width of order the correlation length $\xi < R$.

(iii) The competition between the electrostatic contributions to the formation free energy, coming from protein-protein repulsion, polyelectrolyte self-repulsion, and protein-polyelectrolyte attraction, leads to interesting re-entrant phase behavior as is shown in Figure 4. Plotted are the critical protein concentration for the onset of aggregation of empty and filled capsids as a function of the ambient salt concentration $[c_s]$ for the optimal case $\langle\phi\rangle = \langle\phi\rangle^*$. Assembly of empty capsids characterizes section A of the diagram; Section B is characterized by the absence of aggregates and Section C by filled capsids. The effective border between the regions of filled and empty capsids is determined by a polyelectrolyte desorption condition.

In section II we begin by generalizing the equilibrium assembly model for charged capsid proteins that was used earlier to explain the salt (and temperature) dependence of empty HBV capsid assembly [8] to describe capsid assembly with a polyelectrolyte cargo. Next, in section III, we discuss the free energy for the adsorption of a generic, oppositely charged, flexible polyelectrolyte onto the inner wall of a fully formed capsid shell. In sections IV and V we separately discuss the concentration profiles under conditions of full, respectively, restricted equilibrium of the trapped polyelectrolyte molecules. In section VI, we combine the results of sections II through V to obtain assembly phase diagrams for full and in restricted equilibrium. The implications and limitations of our results are discussed in Section VII where we compare our results with the Bancroft study of co-assembly of CCMV and homopolymeric RNA. We conclude with a brief discussion of key differences between polyelectrolyte encapsidation and viral assembly.

## II. Thermodynamics of Capsid Self-Assembly

In order to substantiate the claims made in the Introduction, we start by defining the model (see Fig. 1). A dilute solution contains both single protein subunits (or oligomers such as dimers or pentamers if that happens to be appropriate), which can aggregate into capsid shells containing *N* subunits each, as well as negatively charged flexible polyelectrolytes. We will assume the solvent conditions to be good. The protein contribution to the free energy density of the solution is then:

$$f/k_B T = c_m \ln \omega c_m - c_m + c_c \ln \Omega c_c - c_c + c_c \Delta G_c + \chi c_m \phi_B. \qquad (1)$$

The first four terms are the ideal solution free energy densities of free and aggregated monomers respectively, with $c_m$ the concentration (number density) of free protein subunits (or oligomers) and $c_c$ that of the capsids; the parameters $\omega$ and $\Omega$ can be viewed as interaction volumes per protein, respectively, per capsid. We assume that only one type of protein subunit and only one type of *N*-subunit capsid are present in the solution. The two concentrations are linked by the condition that the total protein concentration, *c*, is fixed,

$$c_m + N c_c = c. \qquad (2)$$

In the fifth term of Eq. 1,

$$\Delta G_c = \Delta G_0 + \Delta G_p, \qquad (3)$$

represents the formation free energy of a capsid from *N* individual subunits. At the level of a Debye-Hückel (DH) theory of linearized electrostatic interactions, $\Delta G_c$ can be written as the sum of ΔG$_0$, the free energy of formation of capsids in the absence of polyelectrolyte, and a contribution from the polyelectrolyte, ΔG$_p$. As argued elsewhere [8], a plausible form for ΔG$_0$ is

$$\Delta G_0 / N \approx -\gamma_H a_H + \sigma^2 \lambda_B \lambda_D a_C, \tag{4}$$

where $\gamma_H$ is the free energy gain per unit area (in units of $k_BT$) resulting from the removal of the water-exposed hydrophobic patches of the capsid protein subunits from an aqueous to a dielectric environment, and $a_H$ the total hydrophobic area of a single monomer buried during aggregation, of the order 10 nm$^2$ [7]. (See Figure 1.) It provides the principle driving force of (empty) capsid assembly. The hydrophobic interface energy $\gamma_H$ does not depend on salinity, at least to a first approximation, but it is an increasing function of temperature [20].

The second term in Eq. 4 originates from the electrostatic self-repulsion of a uniformly charged shell with surface charge density σ, obtained within the Debye-Hückel approximation. Here, $\lambda_B = e^2/4\pi\varepsilon k_BT$ is the Bjerrum length of water at a temperature $T$, with $e$ the unit charge, ε the dielectric constant of water, $\lambda_D = 1/\sqrt{8\pi\lambda_B N_A [c_s]}$ the Debye screening length, with [$c_s$] the molar concentration of 1-1 salt and $N_A$ Avogadro's number, and finally $a_C$ the charged area of a capsid protein facing the inner surface of the shell (approximately 20 nm$^2$ for a T = 4 virus [8]). For water at room temperature, $\lambda_B \approx 0.7$ nm and $\lambda_D \approx 0.3/\sqrt{[c_s]}$ nm. The Debye screening length $\lambda_D$ is assumed to be small compared with the capsid radius $R$. Equation 4 was used in reference [8] to fit the self-assembly thermodynamics of HBV capsids in solution, allowing the determination of $\sigma$ and $\gamma_H$.

Finally, the last term of Eq. 1 describes the attractive electrostatic interactions between the remaining free protein monomers in solution and the polyelectrolyte material at a monomer concentration $\phi_B$ in terms of a (negative) Flory χ parameter, which is discussed below.

After minimization of the free energy density with respect to the capsid concentration $c_m$, subject to the constraint of Eq. 2, one obtains the law of mass action for the problem in hand:

$$\frac{c_c}{c_m^N} = \frac{\omega^N}{\Omega} \exp{-\Delta \tilde{G}_c} \tag{5}$$

with $\Delta \tilde{G}_c = \Delta G_c - N\chi c_B$. In the relevant limit $N \gg 1$, the fraction of proteins $f = Nc_c/c$ in aggregate form is given by a classical relation of equilibrium self-assembly:

$$f \approx \begin{cases} 1 - c^*/c & c > c^* \\ 0 & c < c^* \end{cases} \tag{6}$$

The threshold concentration $c^*$ may be considered as a *critical subunit concentration*, or CSC, in analogy to the well-known critical micelle concentration for self-assembling surfactant molecules [20]. The CSC is controlled by the capsid free energy according to:

$$c^* \approx \omega^{-1} \exp{\frac{\Delta \tilde{G}_c}{N}}. \tag{7}$$

Using Eq. 4 in Eq. 7 gives the CSC for formation of *empty* capsids if we set $\Delta G_p$ and $\chi$ equal to zero. Equation 7 predicts that - for empty capsids - the CSC rapidly decreases if we increase the concentration of added salt or if we strengthen the apolar character of the hydrophobic patches of the capsid proteins by raising the temperature [8]. Experiments on HBV confirm this conclusion [7].

## III. Polyelectrolyte Encapsidation.

We now specify the nature of the polyelectrolyte cargo of the capsid. The (effective) linear charge density along a flexible polyelectrolyte of $M$ monomers, or Kuhn segments, of length $l$, will be presumed to be of the order $\alpha \approx 1$ per Kuhn length, as for instance appropriate for homo-polymeric ssRNA. Under these conditions, we can ignore the dependence $l$ on the salinity of the solution [21]. The electrostatic self-repulsion of the flexible chains is accounted for by an excluded volume $v$ acting between any two Kuhn segments. Let $\psi(r) = \dfrac{e}{4\pi\varepsilon r}\exp{-r/\lambda_D}$ be the DH point charge potential. The electrostatic contribution to the excluded volume is then [22]

$$v = \int d^3\vec{r}\left(1 - \exp\left(-\alpha^2 \frac{e\psi(r)}{k_B T}\right)\right) \approx 4\pi\alpha^2 \lambda_B \lambda_D^2, \tag{8}$$

assuming that the electrostatic energy per Kuhn segment is less than the thermal energy [23]. Next, the electrostatic interaction between the polyelectrolyte material and the proteins is included via a *surface energy*:

$$F_S / A \approx -\alpha e \int_0^\infty dz\, \psi(z)\phi(z) \tag{9}$$

Here, $\phi(z)$ is the number of Kuhn segments per unit volume at a distance $z$ from the (charged) protein surface, and $\psi(z)$ is the electrostatic potential of a double-layer with (uniform) surface charge density $\sigma$. (See also Figure 1.) Within the DH approximation, for a flat plate [22]:

$$\frac{e\psi(z)}{k_B T} = 4\pi \lambda_D \lambda_B \sigma \exp(-z/\lambda_D), \tag{10}$$

which is appropriate in the thin double-layer limit, $\lambda_D \ll R$. If the polyelectrolyte monomer concentration does not vary significantly over the Debye length, then Eq. 9 reduces to

$$F_S/A \approx -4\pi\alpha\sigma\lambda_B \lambda_D^2 \phi(0) \equiv -\gamma\,\phi(0), \tag{11}$$

Here, $\gamma = 4\pi\alpha\sigma\lambda_B \lambda_D^2$ is a measure of the strength of the surface attraction, while $\phi(0)$ is the monomer concentration at the surface. If we apply Eq. 11 to free protein subunits in the bulk polyelectrolyte solution, then we must equate $\phi(0)$ with the bulk monomer concentration $\phi_B$. Equation 11 then defines an effective Flory parameter $\chi = -\gamma a_C$ for the electrostatic attraction between the proteins and the polymer in the bulk solution, with $a_C$ the appropriate charged area of one subunit (see Eq. 1). If we apply Eq. 11 to a completed capsid then $A$ is the inner capsid surface area ($A \approx N a_C$). In the next two sections, we will apply the methods of polymer physics to obtain $\phi(0)$ for this case.

# IV. Fixed Chemical Potential: Full Equilibrium

In this section we will assume that the polyelectrolyte material inside the capsid is in full chemical equilibrium with a semi-dilute polyelectrolyte bulk solution. Full chemical equilibrium requires the capsid wall to be *permeable* to the polyelectrolyte molecules (see Fig. 1). This limit is not expected to apply to actual viruses, apart from certain special cases such as, perhaps, CCMV capsids at high *pH*, or the "fenestrated" HBV capsids permeable to short RNA sequences. However, the regime of full chemical equilibrium is important for setting a conceptual framework for the next section where we discuss impermeable capsid walls.

Under conditions of full chemical equilibrium, the polyelectrolyte contribution to the formation free energy equals [24]:

$$\Delta \tilde{G}_p = \int_{r<R} d^3\vec{r} \left\{ \frac{1}{6} l^2 (\nabla \phi^{1/2})^2 + \frac{1}{2} v(\phi^2 - \phi_B^2) - \mu_B(\phi - \phi_B) \right\} - 4\pi R^2 \gamma (\phi_S - \phi_B). \quad (12)$$

Here, $\phi(r) = \phi(R-z)$ is the monomer concentration as a function of the radial distance $r = |\vec{r}| = R - z$ from the center of the capsid, so $\phi_S = \phi(r)$ for $r = R$ or $z = 0$ is the surface concentration. Equation 12 is – apart from the last constant – the classical mean-field adsorption free energy of an inhomogeneous, semi-dilute polymer solution in the limit that the number of polymer segments *M* is very large and the "ground-state approximation" holds [24]. The first term describes the entropic free energy cost of an inhomogeneous density profile. The second term is the free energy density due to electrostatic self-repulsion of compressed polyelectrolyte material, expressed in the form of a second virial expression (we subtract the corresponding bulk solution term present before assembly in the same volume). The third term is the chemical work associated with the introduction of the excess polyelectrolyte material from the bulk into the capsid volume; $\mu_B = v\phi_B$ is the polyelectrolyte chemical potential of the bulk solution. The last term describes the electrostatic attraction between the protein capsid and polyelectrolyte material. Again, we subtract the corresponding electrostatic attraction between individual subunits and polyelectrolyte material in solution (the Flory χ term).

Minimization of this free energy is, as usual, carried out conveniently in terms of the "wave function" $\psi(r) \equiv \phi^{1/2}(r)$. It produces a non-linear Euler-Lagrange equation

$$\frac{1}{6} l^2 \nabla^2 \psi = \psi \left( v \psi^2 - \mu_B \right). \tag{13}$$

Outside the capsid, in the uniform bulk solution, $\psi$ equals $\psi_B = \sqrt{\mu_B / v} = \sqrt{\phi_B}$. Linearization of Eq. 13 around this uniform solution value shows that the "healing length" $\xi$, describing the relaxation of deviations from the uniform state is given by

$$\xi^2 \equiv \frac{l^2}{3 v \phi_B}. \tag{14}$$

Physically, $\xi$ corresponds to the "blob size" [24] such that on length scales less than $\xi$, the polyelectrolyte material can be treated as an individual chain, characterized by power-law correlations while correlations are screened on length scales large compared to $\xi$.

Demanding this free energy to be stationary with respect to the surface concentration produces one of the boundary conditions for Eq. 13:

$$\left. \frac{1}{\psi} \frac{d\psi}{dr} \right|_{r=R} = \frac{6\gamma}{l^2} \equiv \frac{1}{d}. \tag{15}$$

Here, $d \equiv l^2 / 6\gamma$ is the extrapolation length mentioned in the Introduction. The other boundary condition states that the concentration at the center of the shell must be a minimum, i.e., $d\psi/dr = 0$ for $r = R$. Inserting Eqs. 13 and 15 in Eq. 12, we can write the polyelectrolyte contribution to the formation energy as:

$$\Delta \widetilde{G}_p = -\frac{1}{2}v \int_{r<R} d^3\vec{r}\left(\psi^4 - \phi_B^2\right) + 4\pi R^2 \gamma \phi_B. \tag{16}$$

It is helpful to consider here the magnitudes of the various length scales: the capsid radius $R$, the correlation length $\xi$, the extrapolation length $d$, the Debye Screening length $\lambda_D$, and the Kuhn length $l$. The screening length is of order 1 nm under standard conditions. For ss RNA or ss DNA homopolymers, the Kuhn length $l$ also is of the order of 1 nm, at least under physiological conditions [25]. For a typical capsid surface with a net surface charge density $\sigma$ of 0.1 – 1 charges per nm$^2$, $d$ is of order 0.1 nm. As discussed in more detail in section VII, this small value of $d$ makes the theory qualitative even for homopolymeric polynucleotides, but should arguably provide a more precise description for other, less strongly charged polyelectrolyte cargo. The capsid radius $R$ on the other hand is much larger, of order 10 – 30 nm. Finally, the correlation length $\xi$ is in determined by the bulk polyelectrolyte concentration $\phi_B$. Depending on $\phi_B$, it could vary from the Kuhn length $l$ at high bulk concentrations to the radius of gyration of a single chain at the lowest concentrations. For a homopolymeric ss RNA chain of 2–4 kilobases the radius of gyration would exceed the radius of a T = 3 capsid. We will be interested in the regime of low bulk concentrations, so we will focus on the case that $\xi$ is large compared the extrapolation length $d$ (i.e., the "strong-adsorption limit") though not necessarily larger than the capsid radius.

We now turn to the solution of Eq. 13. Two regimes of interest, that we will denote as the "exponential", respectively, the "power-law" domain, are determined by the ratio of the correlation length and the capsid radius.

i) *Exponential domain*: $d \ll \xi < R$.

In this regime, the concentration profile consists of a central core region, where the monomer concentration equals the bulk concentration, surrounded by a layer of enhanced concentration covering the interior surface of the capsid. In terms of the distance $z = R - r$ from the capsid surface, the solution of Eq. 13 can be approximated by [26-28]

$$\psi(z) \approx \psi_B \coth\left(\frac{z+d}{\xi}\right), \tag{17}$$

for $z \ll R$. This is the classical concentration profile of a semi-dilute solution of polymers surface-adsorbed on a flat plate (systematic corrections for curvature can be included but do not significantly affect our results [28]). Near the wall, the polyelectrolyte concentration exhibits a power-law divergence for $z + d$ values small compared to the correlation length $\xi$,

$$\phi(z) = \psi(z)^2 \approx \frac{\phi_S}{(1+z/d)^2}. \tag{18}$$

Here,

$$\phi_S = \phi(0) = \frac{l^2}{3vd^2} \tag{19}$$

is the polyelectrolyte concentration at the capsid surface. Note that for $l < d$, this surface concentration approaches $1/v$, the density of a melt, which means that our virial expansion becomes inaccurate. The degree of surface enhancement $\phi_S/\phi_B = \xi^2/d^2$ is determined by the ratio of the correlation length $\xi$ and the extrapolation length $d$.

In the opposite limit, with $z + d$ large compared to $\xi$, $\phi(z)$ approaches exponentially the bulk concentration $\phi_B$ consistent with the role of $\xi$ as a healing length [24],

$$\phi(z)/\phi_B = 1 + 4\exp{-2z/\xi} \qquad (20)$$

for $z = R - r \gg \xi - d$. See also Figure 2. The total surface charge $\sigma_p$ per unit area of excess absorbed polyelectrolyte material equals

$$\sigma_p \approx \alpha \int_0^\infty dz(\phi(z) - \phi_B) \approx \alpha\phi_S d = 2\sigma, \qquad (21)$$

using (in the last two steps) Eqs. 8, 12, and 17. According to Eq. 21, polyelectrolyte adsorption effectively produces a *charge-reversal* of the adsorbing surface, a well-known result from polymer physics [22]. Alternatively, define $\langle\phi\rangle$ to be the mean excess polyelectrolyte concentration inside the capsid. From Eq. 21, it then follows that

$$\langle\phi\rangle \approx \frac{3}{R}\int_0^\infty dz\,\phi(z) \approx \frac{3\phi_S d}{R} \approx \frac{l^2}{vRd} \approx \frac{6\sigma}{R\alpha}, \qquad (22)$$

provided again that the correlation length $\xi$ is small compared to $R$. Remarkably, the mean excess polyelectrolyte concentration depends only on the surface charge density and the radius of the capsids, and is independent of the salt concentration. This would not have been the case if we had fixed the electrical surface *potential* instead of the surface *charge* [29].

ii) *Power-law domain*: $\xi > R \gg d$.

If we lower the bulk concentrations to the point that the correlation length exceeds the capsid radius, then the central core disappears. The power-law behavior Eq. 18 previously confined to the vicinity of the surface now extends throughout the capsid; see Figure 2. Using a series expansion solution of Eq. 13, it is easy to show that for small $r$, i.e., near the center of the capsid, the solution must have the form:

$$\psi(r) = \psi(0) + \frac{v}{l^2}\psi(0)\left(\psi(0)^2 - \phi_B\right)r^2 + O(r^4). \tag{23}$$

On the other hand, near the inner surface of the capsid at $r = R$, we must recover a power-law divergence of the form of Eq. 18. Specifically, we demand that $\psi(r) \propto 1/(r_0 - r)$ with $r_0$ some constant. A trial function that is accurate in both small and large $r$ regimes is then:

$$\psi(r) = \frac{\Xi}{1 - r^2/r_0^2} \tag{24}$$

with $\Xi$ and $r_0$ constants to be determined. These we fix by considering the behavior of our Ansatz Eq. 24 near $r = 0$ and near $r = R$. From the boundary condition Eq. 15, it follows that

$$R/r_0 \approx 1 - \delta \tag{25}$$

with $\delta = d/R \ll 1$. From the small $r$ expansion, Eq. 23, we find

$$\Xi^2 \approx \phi_B\left(1 + 3\frac{\xi^2}{R^2}\right). \tag{26}$$

Note that the concentration at the center of the capsid now exceeds the bulk concentration. Note further that the surface concentration obeys

$$\phi_S = \frac{\Xi^2}{\left(1 - R^2/r_0^2\right)^2} \approx \frac{\Xi^2}{4\delta^2}. \tag{27}$$

In the limit $\xi/R \gg 1$, this reduces to $\phi_S/\phi_B \approx (3/4)\xi^2/d^2$, which means that Eq. 19 for the surface concentration approximately holds in both regimes.

Having obtained approximate solutions to Eq. 13 for the regimes of interest, we can compute the polyelectrolyte contribution to the capsid formation energy using Eq. 16. The functional form of the polyelectrolyte contribution to the formation energy is, to leading order, the same in the two regimes:

$$\Delta \widetilde{G}_p \approx -Ad \frac{v\phi_S^2}{6}\left(1 + O\left(\frac{d^2}{\xi^2}\right)\right), \tag{28}$$

with the surface concentrations given by Eq. 19, respectively, Eq. 27. The formation energy is thus directly proportional to surface area $A$ of the capsid. Note that the term $v\phi_S^2$ has the form of the second viral free energy density in the surface layer with an effective thickness d *except* that the sign is negative. The negative sign means that, under conditions of full chemical equilibrium, the presence of the polyelectrolyte material promotes the formation of capsids.

# V. Fixed Packing Density: Restricted Equilibrium

We now turn to the case where capsids that are *not* permeable to the polyelectrolyte material. *How* the polyelectrolyte material is captured, i.e., the assembly pathway, will be left outside our considerations for reasons discussed in the Introduction. We will focus entirely on the question whether or not polyelectrolyte encapsidation lowers the free energy of the system. We will assume here encapsidation of just a single, long polyelectrolyte molecule (in the Conclusion, we will discuss the case of multiple captures). The number of monomers $M$ is assumed sufficiently large so the radius of gyration of the polymer is large compared with the capsid radius $R$. The polyelectrolyte material trapped inside the capsid will again be treated as a semi-dilute solution but the correlation length is no longer pre-determined. Instead of the chemical potential, we now must fix the mean concentration $\langle \phi \rangle = M/V$, with $V$ the capsid volume. Because the shell is now impermeable to the polyelectrolyte, but *not* to the solvent, we must expect there in general to be an *osmotic pressure* difference $\Delta \Pi$ across the capsid wall. Similarly, since capsid walls are permeable to small ions, there also should be an *electrical potential* difference of the Donnan type across the capsid wall.

The work required to compress the polyelectrolyte molecule into the capsid volume – in other words: the change in free energy – now equals

$$\Delta \widetilde{G}_p = \int_{r<R} d^3\vec{r} \left\{ \frac{1}{6} l^2 \left( \nabla \phi^{1/2} \right)^2 + \frac{1}{2} v \phi^2 \right\} - 4\pi R^2 \gamma \phi_S. \tag{29}$$

We actually still should have subtracted a term corresponding to the free energy of the polyelectrolyte molecule in free solution but under the stated conditions this term is negligible. We must minimize Eq. 29 subject to the condition that the mean concentration $\langle \phi \rangle = M/V$ is fixed. The resulting Euler-Lagrange equation for the concentration profile is again Eq. 13, in terms of the wave function $\psi = \phi^{1/2}$, except that the chemical potential $\mu_B$ now must be viewed as Lagrange multiplier. To avoid confusion, we will denote the Lagrange multiplier by $\mu$. The approximate solutions we found in the last section carry over to the present case and we again must distinguish the exponential and power-law regimes.

   i) *Exponential domain*: $d \ll \xi < R$.

First assume that the polyelectrolyte material inside the capsid is characterized by the profile Eq. 17, with $\phi_B$ replaced by $\mu/v$. The concentration is again uniform in the center of the capsid. The Lagrange multiplier is fixed by the condition of mass conservation,

$$\langle \phi \rangle \approx \frac{\mu}{v} + \frac{3}{R}\int_0^\infty dz\left(\psi(z)^2 - \frac{\mu}{v}\right). \tag{30}$$

Inserting Eq. 17 in Eq. 30 gives

$$\langle \phi \rangle \approx \frac{\mu}{v} + 3\frac{d}{R}\phi_S, \tag{31}$$

with $\phi_S$ the surface concentration Eq. 19. Next, inserting Eq. 17 in the polyelectrolyte contribution to the formation free energy, and eliminating the Lagrange multiplier, using Eq. 31, gives

$$\Delta \tilde{G}_p(\langle \phi \rangle) \approx -Ad\frac{v\phi_S^2}{6} + V\frac{v}{2}\left(\langle \phi \rangle - 3\frac{d}{R}\phi_S\right)^2 \tag{32}$$

where we require that $\langle \phi \rangle > \langle \phi \rangle^* \equiv 3d\phi_S/R$. Note the separation in surface and volume contributions (*A* is again the capsid surface area and *V* the capsid volume). The surface term is, to leading order, equal to the formation free energy at fixed chemical potential (see Eq. 28). The volume term represents the energy cost produced by the electrostatic self-repulsion of the excess polyelectrolyte cargo when $\langle \phi \rangle$ exceeds $\langle \phi \rangle^*$. The condition $\langle \phi \rangle > \langle \phi \rangle^*$ follows because the correlation length,

$$\xi = \frac{l}{\sqrt{3\mu}} = \frac{l}{\sqrt{3v\left(\langle\phi\rangle - \langle\phi\rangle^*\right)}}, \tag{33}$$

has to remain less than the capsid radius $R$ if we want the exponential regime to be valid.

As a function of $\langle\phi\rangle$, the free energy Eq. 32 has a minimum at $\langle\phi\rangle = \langle\phi\rangle^*$ right at the border of the range of validity of the exponential regime. On the other hand, the *maximum* packing density $\langle\phi\rangle_{max}$ that possibly can be achieved by self-assembly is defined by the condition that the attractive protein-polymer attraction exactly balances the polymer self-repulsion in the core region, i.e., that $\Delta\tilde{G}_p = 0$. This happens when:

$$\langle\phi\rangle_{max} = \left(1 + \frac{1}{3}\sqrt{\frac{R}{d}}\right)\langle\phi\rangle^*. \tag{34}$$

This maximum packing density considerably exceeds the optimal packing density by a factor of order $\sqrt{R/d} \gg 1$, though it still is much less than the surface concentration $\phi_S \approx \frac{1}{3}\langle\phi\rangle_{max}\sqrt{R/d}$. The minimal healing length at this maximum packing density equals $\xi_{min} = d(R/d)^{1/4}$.

As noted already, an osmotic pressure $\Delta\Pi \equiv -d\Delta\tilde{G}_p / dV\big|_{M,N}$ is exerted on the capsid wall under conditions of restricted equilibrium. When taking here the derivative of the free energy with respect to the volume, it is important to keep both the number of enclosed polyelectrolyte monomers $M$ and the number of capsid subunits $N$ fixed. For instance, the mean packing density $\langle\phi\rangle \propto 1/V \propto 1/R^3$ is inversely proportional to $V$ under these conditions. Since the total capsid charge $A\sigma$ — with $\sigma$ the capsid surface charge density — is fixed it follows that $\sigma \propto 1/A$ so that the extrapolation length $d \propto 1/\sigma \propto A \propto R^2$. Hence, the surface density drops with $R$ as $\phi_S \propto \frac{1}{d^2} \propto \frac{1}{R^4}$. Keeping these conditions in mind, we find

$$\frac{\Delta \Pi}{k_B T} \approx \frac{1}{2} v \left( \langle \phi \rangle - 3 \frac{d}{R} \phi_S \right)^2 - \frac{2}{3} v \phi_S^2 \frac{d}{R} \qquad (35)$$

In order to interpret this result, we note that osmotic pressure inside a spherical container produces a tension $\tau = \Delta \Pi R / 2$ on the wall. In our case, this tension must be absorbed by the interaction potential that holds the subunits together (e.g., the electrostatic and hydrophobic forces discussed in Section II). The first term of Eq. 35 is the osmotic pressure of the uniform core region where the monomer concentration equals $\langle \phi \rangle - 3 \frac{d}{R} \phi_S$. The second term is a contribution coming from the surface layer with enhanced monomer concentration (see Eq. 27). Since the wall tension equals $\tau = \Delta \Pi R / 2$ and since this term is inversely proportional to $R$, we can interpret it as an effective *negative* wall tension $\Delta \tau = -\frac{1}{3} k_B T v d \phi_S^2$, which must be absorbed by the bending rigidity of the capsid wall. The physical reason for this negative tension is that by reducing the wall surface area at a fixed number of surface charges, the polymer/capsid binding energy is increased. If we divide the polyelectrolyte material between a "surface" and a "bulk" part, and consider the capsid wall plus the surface-adsorbed part as constituting an effective interface, then we can view the first term of Eq. 35 as a positive osmotic pressure exerted on this effective interface. However, even at the maximum packing density, the *total* osmotic pressure

$$\frac{\Delta \Pi_{max}}{k_B T} \approx -\frac{1}{6} v \phi_S^2 \frac{d}{R} \qquad (36)$$

still is negative and of the form of a negative contribution to the wall tension. Self-assembly is apparently not able to "load" a capsid to the point that the capsid wall is under a net positive tension.

ii) *The power-law domain*: $\xi > R \gg d$.

If we reduce the length of the captured polyelectrolyte molecule, we enter the power-law regime, where we must use Eq. 24 in Eq. 29. The Lagrange multiplier is fixed by the condition that the average concentration must equal $\langle \phi \rangle$:

$$\langle \phi \rangle = \frac{3}{R^3} \int_0^R dr\, r^2 \psi(r)^2 \approx \frac{3\Xi^2}{4\delta}. \tag{37}$$

Here, $\delta = d/R$, which again follows from the boundary condition Eq. 15. The formation energy becomes

$$\Delta \widetilde{G}_p \approx -\frac{4\pi}{27} l^2 \langle \phi \rangle \frac{R^3}{d^2} + \frac{2\pi}{27} v \langle \phi \rangle^2 \frac{R^4}{d} \tag{38}$$

The validity condition for Eq. 38 is that $\xi > R$, or, equivalently, that $\langle \phi \rangle < \langle \phi \rangle^* = 3d\phi_S/R$. Note that the dependence on the capsid radius $R$ no longer separates into surface and volume terms, which is due to the extended, power-law density profile. This form of $\Delta \widetilde{G}_p$ has a minimum near $\langle \phi \rangle = \langle \phi \rangle^*$, i.e., at the border of the validity range of the power-law regime. Comparing Eq. 38 with Eq. 32, we find that the free energy gain of encapsidation is maximized at the crossover point between the exponential and power-law regimes, i.e., *when the correlation length $\xi$ is comparable to the radius R of the capsid* (see Figure 3). Note that the formation free energy at the minimum, $\Delta \widetilde{G}_p \approx -Adv\phi_S^2/6$, where Eqs. 32 and 38 match, coincides with the capsid formation free energy under conditions of full chemical equilibrium.

A physical explanation for the result that $\xi \approx R$ for capsids with maximum thermodynamic stability is as follows. When $\xi > R$, the concentration profile of the polymeric segments in the vicinity of the surface is sub-optimal since we can continue to lower the surface energy by adding polyelectrolyte material to the capsid interior.

On the other hand, for $\xi < R$ the surface concentration profile *is* optimal but the excess polyelectrolyte material in the core of the capsid increases the free energy due to electrostatic self-repulsion. When $\xi \approx R$, the concentration profile is optimal with no excess polyelectrolyte material in the core. The required number $M^*$ of monomers for optimal capsid stability is proportional to the capsid surface area:

$$M^* \approx \frac{4\pi}{3} R^3 \langle \phi \rangle^* \approx \frac{4\pi}{3} \frac{R^2 l^2}{vd}. \tag{39a}$$

By comparison, the maximum number $M_{max}$ of monomers scales as

$$M_{max} \approx \frac{4\pi}{3} R^3 \langle \phi \rangle_{max} \approx \frac{4\pi}{9} \frac{R^{5/2} l^2}{v d^{3/2}}. \tag{39b}$$

# VI. Capsid Assembly Diagrams

We now return to question raised in Section II: what is the critical subunit concentration CSC, denoted $c^*$, for the self-assembly of a capsid with a polyelectrolyte cargo? Recall that $c^*$ depends on the formation free energy per subunit as:

$$c^* \approx \omega^{-1} \exp\frac{\Delta \tilde{G}_c}{N} = \omega^{-1} \exp\frac{\Delta G_0 + \Delta \tilde{G}_p}{N}. \qquad (7')$$

First consider that case that self-assembly takes place under optimal conditions, i.e., under conditions of full chemical equilibrium. The filling concentration $\langle \phi \rangle^*$ is in this case fixed at:

$$\langle \phi \rangle^* = \frac{6\sigma}{R\alpha}, \qquad (40)$$

expressed in terms of the surface charge density. We can use Eq. 28 for both exponential and the power-law regimes in Eq. 7:

$$\ln \omega c^* \approx -\gamma_H a_H + \sigma^2 \lambda_B \lambda_D a_C \left\{ 1 - 4(4\pi)^2 \alpha \sigma l^{-2} \lambda_B \lambda_D^3 \right\} \qquad (41)$$

The first two terms of Eq. 41 together equal the formation energy per protein (in units of $k_B T$) for an empty capsid (see Section II), while the third term represents the increased formation energy of a filled capsid (see Eq. 28). Note the *cubic* dependence on the surface charge density. The two electrostatic contributions to Eq. 41 have a different dependence on the salt concentration $[c_s]$: the repulsive term scales as $\lambda_D \propto [c_s]^{-1/2}$ while the attractive term scales as $\lambda_D^4 \propto [c_s]^{-2}$. It follows that at low ionic strengths the attractive polyelectrolyte/capsid interaction dominates and at high ionic strength the subunit self-repulsion.

Figure 4 shows the critical subunit concentrations of both empty and filled capsids. For high salt concentrations, the two critical concentrations nearly coincide. If the subunit concentration is increased under those conditions, both empty and filled capsids will start to assemble at about the same protein concentration. On the other hand, at low salt concentrations, the critical subunit concentration of filled capsids is significantly lower than that of empty capsids. Now, mostly filled capsids form if one raises the protein concentration. Clearly, it is this second regime that would be the relevant one for (synthetic) viral self-assembly. As shown in Figure 4, the two regimes are separated by a maximum of the critical subunit concentration as a function of the salt concentration. The condition for the maximum is that the surface charge density equals:

$$\sigma^* = \frac{l^2}{\alpha(16\pi)^2 \lambda_B \lambda_D^3}. \tag{42}$$

The surface charge density of a capsid has to exceed $\sigma^*$ for self-assembly to produce (mostly) filled capsids. The corresponding critical subunit concentration equals:

$$\ln \omega c_{max} \approx -\gamma_H a_H + \frac{3}{4}\sigma^2 \lambda_B \lambda_D a_C. \tag{43}$$

A physical interpretation of Eq. 42 is obtained by applying the theory of polymer desorption. A single polyelectrolyte molecule in the neighborhood of an oppositely charged surface undergoes a *desorption transition* when the conformational chain entropy exceeds the opposing enthalpic binding energy. The condition for the desorption transition has the same form as Eq. 42 [30]. In other words, a condition for the self-assembly process to produce filled caspids is that the subunit surface charge must be sufficiently high for the polyelectrolyte molecules to adhere to the inner surface of the assembling capsid.

We now turn to the case of restricted equilibrium with $\langle\phi\rangle > \langle\phi\rangle*$. Using Eq. 32 in Eq. 7 gives:

$$\ln\omega c^* \approx -\gamma_H a_H + \sigma^2 \lambda_B \lambda_D a_C \left\{1 - 4(4\pi)^2 \alpha\sigma l^{-2} \lambda_B \lambda_D^3 \right\} \\ + \frac{2\pi}{3}\alpha^2 \lambda_B \lambda_D^2 a_C R(\langle\phi\rangle - \langle\phi\rangle*)^2. \tag{44}$$

The new term is due to the electrostatic self-repulsion of the excess polyelectrolyte material in the core of the capsid. The new term depends on the salt concentration as $\lambda_D^2 \propto [c_s]^{-1}$, which is *intermediate* between the salt-dependence of the subunit-subunit repulsion and that of the subunit-polyelectrolyte attraction. This has important consequences. Start at very high salt concentrations (see Fig. 4b). As before, $c^*$ is at first dominated by subunit self-repulsion so the CSC for assembly of empty and filled capsids is about the same. Now start to lower the salt concentration. The electrostatic self-repulsion of the core first grows in strength with the result that the CSC for filled capsids *exceeds* that of empty capsids. In other words, empty capsids form in preference over filled capsids as we raise the subunit concentration. Further lowering of the salt concentration causes the subunit-polyelectrolyte interaction to grow in strength, which causes a drop in $c^*$. The CSC for empty and filled capsids is equal when the filling concentration equals the maximum concentration Eq. 35, i.e., when

$$\langle\phi\rangle = \frac{4\sigma^{3/2}}{\alpha l}\sqrt{\frac{6\pi\alpha\lambda_B\lambda_D^2}{R}}, \tag{35'}$$

in terms of the original quantities. For even lower salt concentrations, the filling concentration is less than $\langle\phi\rangle_{max}$ and filled capsids form in preference over empty capsids (at least under quasi-equilibrium conditions).

This result is interesting because it means that under condition of restricted equilibrium, we should expect a rather sharp transition, as a function of salt concentration, between the assembly of filled and empty capsids namely when Eq. 35' is satisfied. Recall that under conditions of full chemical equilibrium, empty and filled capsids both form at higher salt concentrations.

# VII. Conclusion

In this conclusion we will re-examine some key results of our study, discuss their implications for self-assembly studies of synthetic viruses such as the Bancroft et al. study [4], and then compare our results with what is known about actual viruses.

We found that under conditions of full chemical equilibrium, the monomer filling concentration should equal $\langle\phi\rangle^* = \frac{6\sigma}{R\alpha}$. The fact that $\langle\phi\rangle^* = \frac{6\sigma}{R\alpha}$ decreases as the radius increases, is reasonable since capsid filling is driven by a surface energy. The relevance of this result to actual experiments could be questioned since it was derived under conditions of full chemical equilibrium. We saw that under the arguably more realistic conditions of restricted equilibrium, a much larger range of filling concentrations was possible in principle. In this conclusion, we will argue first why we believe that $\langle\phi\rangle^* = \frac{6\sigma}{R\alpha}$ still effectively determines the filling fraction of synthetic viruses, i.e., viral capsids with a polyelectrolyte cargo.

Assume capsid walls that are impermeable to the polyelectrolyte cargo. In that case, $\langle\phi\rangle^* = \frac{6\sigma}{R\alpha}$ represents the *minimum* of the capsid free energy $\Delta\tilde{G}_p(\langle\phi\rangle)$. Assume first that an assembling capsid contains a single polyelectrolyte molecule, as we did in the previous sections. The condition $\langle\phi\rangle^* = \frac{6\sigma}{R\alpha}$ would then be obeyed only if the number of monomers, $M$, equaled $M^* = \frac{8\pi\sigma R^2}{\alpha}$. Assume that $M$ is less than $M^*$ so the filling fraction will drop below $\langle\phi\rangle^* = \frac{6\sigma}{R\alpha}$. If $M$ was equal to $1/2M^*$, then $\Delta\tilde{G}_p(\langle\phi\rangle)$ could be minimized by introducing *two* molecules, each having $1/2M^*$ monomers, into a partially formed capsid. In fact, for any $M$ less than $1/2M^*$, we always can obtain capsid formation energies in the range between $\Delta\tilde{G}_p(\langle\phi\rangle^*)$ to $\Delta\tilde{G}_p(\tfrac{1}{2}\langle\phi\rangle^*)$ by introducing a suitable number of cargo molecules into the capsid.

We infer that, provided the capsid assembly process proceeds sufficiently slowly, the filling fraction should be in the range $\frac{3\sigma}{R\alpha} \leq \langle\phi\rangle \leq \frac{6\sigma}{R\alpha}$ if $M$ is less than $M^*$ [31].

What if $M$ exceeds $M^*$? Consider the following thought experiment. Assume that during assembly, a partially formed capsid shell contains a (small) aperture through which polyelectrolyte molecules with $M > M^*$ can enter or leave the capsid interior. Starting from $\langle\phi\rangle = 0$, the capsid free energy $\Delta\tilde{G}_p(\langle\phi\rangle)$ will decrease as $\langle\phi\rangle$ increases, which means that polyelectrolyte insertion proceeds spontaneously. This will continue until $M^*$ monomers have been introduced and $\langle\phi\rangle = \frac{6\sigma}{R\alpha}$. Increasing $M$ beyond $M^*$ would demand *increasing* $\Delta\tilde{G}_p(\langle\phi\rangle)$. Insertion would *not* proceed spontaneously (unless some external source of thermodynamic work was available such as the DNA insertion motor protein of the $\Phi 29$ bacteriophages [32]). The typical amount of work $Vv\langle\phi\rangle^2 k_B T$ required to squeeze the remaining material into the capsid (of order $10^2 k_B T$ for the estimated values given below) is much too large for a random thermal fluctuation to be able to complete the insertion process (after which the aperture could be closed). Self-assembly of a capsid in the presence of polyelectrolytes with $M > M^*$ is thus expected to produce a *defected* capsid with $M^*$ monomers inside the capsid, and a random coil of $M$–$M^*$ monomers remaining as a tether outside the partially formed shell. We conclude that *well-formed* self-assembled synthetic capsids are expected to have filling concentrations in the range $\frac{3\sigma}{R\alpha} \leq \langle\phi\rangle \leq \frac{6\sigma}{R\alpha}$.

How does this compare with the results of the classical studies of Bancroft and co-workers on the self-assembly of CCMV capsid proteins in the presence of homopolymeric single-stranded RNA molecules [4]? We saw that under good solvent conditions, in the physiological range, the Debye screening length is of order 1 nm, the Kuhn length (of homopolymeric ss RNA) of order 1–2 nm, and the charge parameter $\alpha \approx 1$ [25]. The excluded volume parameter $v$ is then of order 7 nm$^3$ according to Eq. 8. The inner radius $R$ of a CCMV capsid is about 10 nm, as for most T = 3 viruses. CCMV capsid proteins have a nominal positive charge of +10 on their inner surface [10], which translates to a surface charge density $\sigma$ of about 1.4 charges per nm$^2$. The extrapolation length $d$ is then of order 0.1 nm.

The predicted optimal monomer filling concentration $\langle \phi \rangle = \frac{6\sigma}{R\alpha}$ is of order 8 per nm$^3$, corresponding to about 3,500 RNA bases in total. In the Bancroft study [4], homopolymeric ss RNA molecules were used of various lengths, mostly of the order of 500 bases. It was found, from sedimentation experiments, that the *maximum* amount of encapsidated homopolymer RNA material had a molecular weight of about half the actual CCMV RNA genome, which corresponds to about 1,500 bases. The theory significantly overestimates the maximum amount of polyelectrolyte material that can be encapsidated by a synthetic virus.

The obvious origin for the discrepancy is the fact that the (estimated) extrapolation length $d$ was significantly less than not only the (estimated) Kuhn length $l$ but also less than the Debye length $\lambda_D$, except at exceedingly elevated ionic strengths. This implies – see Eq. 19 – that the monomer concentration at the surface would significantly exceed the excluded volume density $1/v$. At the very least, higher-order virial terms would have to be included in the surface layer, which would limit the density and reduce the amount of encapsidated material. In fact, the strength of the electrostatic interaction between the RNA molecules and the CCMV capsid proteins in the surface layer appears to be so large that physical phenomena are to be expected that lie beyond the range of the classical poly-electrolyte physics as discussed in this paper. One possibility is the formation of surface arrays of double-stranded RNA material compensating the capsid surface charge as reported for many T = 3 viruses [33] or Wigner crystallization phenomena [34].

A quantitative test of the theory presented in this paper would be to repeat the Bancroft study but using, instead of ss RNA, polyelectrolyte molecules with α values less than 0.1 to reduce the strength of the interaction. In fact, certain derivatives of polystyrene sulfonate have an *adjustable* value of α [35], which would be a very interesting material as a test cargo. Alternatively, *mutants* of the CCMV subunit are available with a smaller number of positive charges. Such studies would allow a test of the prediction that the maximum amount of encapsidated homopolymeric RNA material should be proportional to the surface charge and inversely proportional to α. A second important test of the theory concerns the *charge reversal* phenomenon. The macro-ion charge of a capsid with $\langle\phi\rangle = \frac{6\sigma}{R\alpha}$ should be opposite in sign to that of the empty capsid. This could be tested in an electrophoresis study: empty and filled capsids should drift with similar speed but in opposite directions. For sub-optimal capsids with $\frac{3\sigma}{R\alpha} < \langle\phi\rangle < \frac{6\sigma}{R\alpha}$ – which can be separated out through their lower sedimentation rates – the macro-ion charge will be proportionally smaller but the *sign* of the charge always should be opposite to that of the empty capsid.

A prediction that may be harder to test concerns the density profile of the monomers inside the capsid, which is claimed to have the form $\phi(r) = \Xi^2 / (1 - r^2 / r_0^2)^2$ under the conditions of slow assembly discussed above. RNA density profiles of natural viruses have been obtained by a combination of X-ray diffraction and Cryo-TEM and it would be extremely interesting if these experiments could be repeated for self-assembly of CCMV proteins with homopolymeric RNA or other synthetic polyelectrolytes [32].

Our second important result concerns the requirement that the capsid surface charge must exceed the desorption threshold $\sigma^* = \frac{l^2}{\alpha(16\pi)^2 \lambda_B \lambda_D^3}$ (see also Fig. 4). Using our earlier estimates for the CCMV-homopolymeric RNA system under physiological conditions, we find that $\sigma^*$ is less than $10^{-3}$ nm$^{-2}$, i.e., *much smaller* than the typical surface charge of a T = 3 viral subunit. This would indicate that CCMV-RNA aggregates are far below the desorption limit and thus very stable. In fact, the surface charge would remain significantly larger than $\sigma^*$ *even if we increased the salt concentration from 0.1 M to 1 M*, a conclusion that must be greeted with some skepticism. Although reassembly of CCMV with viral RNA is efficient at lower salinity levels, it does not take place at 1 M salt (at a *pH* of 7.4). The salt dependence of reassembly of CCMV with homopolymeric RNA has not been reported on, but we suspect that the predicted desorption threshold may be an overestimate of capsid stability for the same reasons that we overestimated the filling fraction: excluded volume effects will limit the concentration of polyelectrolyte material adjacent to the capsid surface. A measurement of the desorption threshold as a function of salt concentration clearly would be a very valuable test of the proposed theory.

It is interesting to compare the predictions of the model system discussed in the paper with the self-assembly of natural viruses. We saw, for the CCMV case, that a natural T = 3 virus can package about double the amount as the synthetic virus with a homopolymer RNA cargo. It is in fact not very surprising that it is easier to compactify a viral RNA molecule with a complex, hydrogen-bonded secondary and tertiary structure, then a homopolymeric RNA, of the same length, with an open random coil structure. Since the self-interaction of viral RNA effectively lowers the solvent quality, it would be very interesting to repeat the studies of Bancroft and co-workers of RNA homopolymer condensation under reduced solvent conditions, for instance by adding $Mg^{++}$ ions, which tend to condense ss RNA, or by using instead of ss RNA various polystyrene sulfonate derivatives with hydrophobic components [35], and see whether the filling fraction could be increased this way. The work required to compactify RNA would progressively drop as we approached the so-called $\Theta$ point, where the coefficient of the second virial expansion goes to zero.

A second issue concerns the RNA density profile inside the virus. For natural CCMV viruses, the packaging density profile is roughly constant – and comparable to hydrated RNA crystals – except for a central core region, which seems empty [36]. Cryo-EM images of E. Coli-expressed HBV also point at an adsorption layer, presumably containing short E. Coli RNA fragments, of about 3 nm width and roughly constant density, and a centre that appears empty [37]. This does not resemble the power-law density profile of the present theory. We believe that it is possible that the density profile under reduced solvent conditions, i.e. near the $\Theta$ point, would involve a boundary surface separating a high and a low density region. The reported density profile of ss RNA viruses also is reminiscent of the spool-like structure of the genome phages [38], which is suggestive that there is some form of liquid-crystalline order in the surface layer of ss RNA viruses.

Another challenging problem that is raised by the comparison with natural viruses concerns the *osmotic pressure*. The arguments presented in this paper indicate that homopolymer encapsidation will produce capsids that are *not* under osmotic pressure, even in the restricted sense of an osmotic pressure exerted by the core region on an effective interface of capsid proteins plus adsorbed polyelectrolyte. This appears logical since it is hard to see how a self-assembling system can produce a pressurized capsid. However, the fact that the genome density inside ss RNA viruses is significantly higher than that of the same molecules in solutions indicates that self-assembling ss RNA viruses in fact may be under internal osmotic pressure. (See, e.g., [39] and works cited therein.) Studies of the genome release scenarios of the FHV and Tymo viruses provide at least qualitative evidence that this internal pressure may play an important role during genome release [12]. Even though a core osmotic pressure is thermodynamically possible, we saw that it is difficult for capsid assembly to be completed when the internal pressure begins to rise during assembly. We speculate that the tertiary structure of viral RNA in solution presents a "condensation surface" for the capsid proteins and that during assembly, as an increasing part of the condensation surface is covered by the growing capsid, the collective self-interaction between different parts of the RNA molecule(s) preventing escape of part of the RNA out of a partially formed capsid, particularly if assembly proceeded at a higher rate. In that case, an interior pressure may be generated. (See also [40].) It would be fascinating if a "biomimetic" protein-polyelectrolyte system could be synthesized that would be able to duplicate this remarkable feat, which seems to be so easy for natural viruses.

*Acknowledgements*: we would like to thank W. Gelbart, C. Knobler, L. Lavelle, Yufang Hu, W. Kegel, and R. Zandi for numerous fruitful discussions. RB would like to thank the NSF for their support under Grant DMR-0404507.

**Figures and Captions**

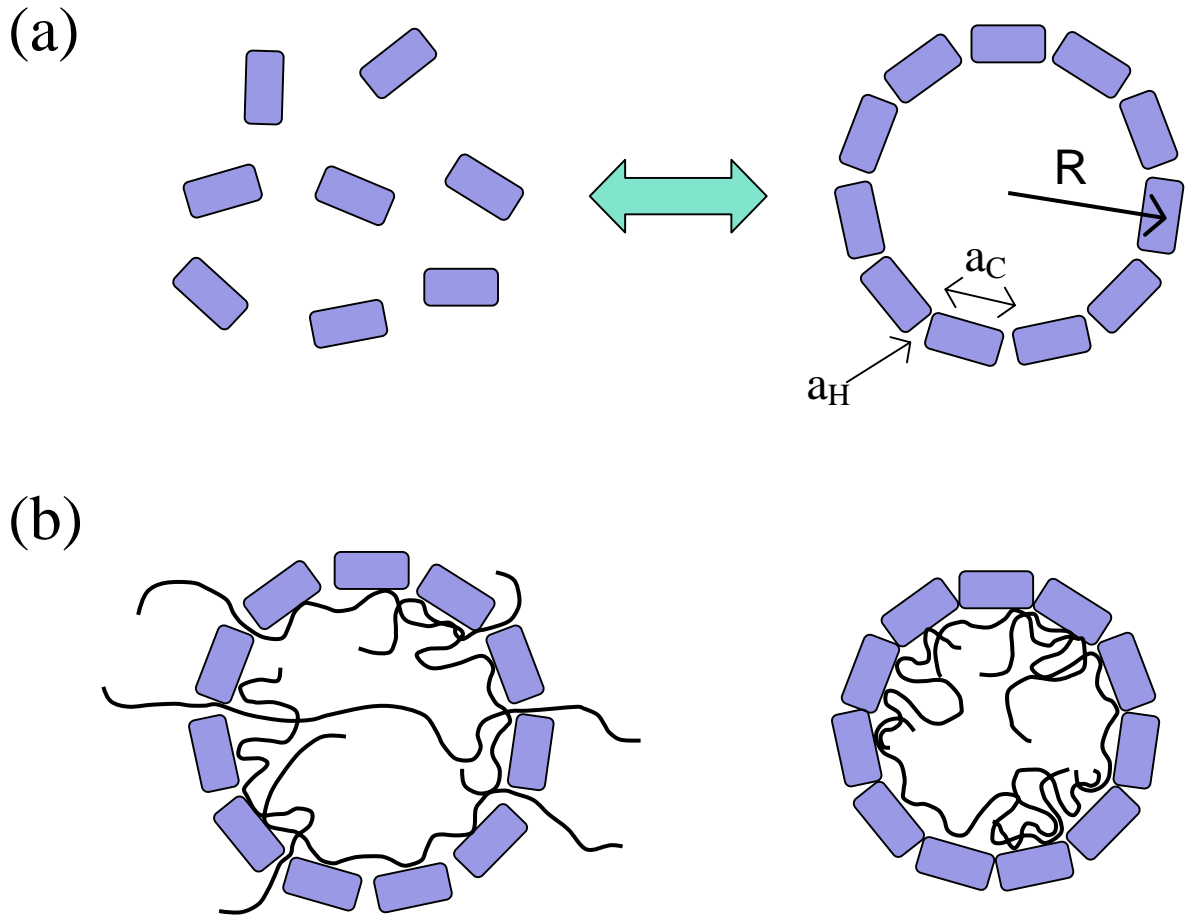

**Figure 1.** (a) Schematic model of capsid self-assembly: protein monomers (left) are in equilibrium with fully formed virus capsids of fixed radius *R* (right). Upon assembly, hydrophobic patches of total area $a_H$ per monomer are shielded from contact with water. The assembly process brings together the charged surfaces of the proteins, with a total area of $a_C$ per protein. (b) In swollen capsids, or in capsids with holes, polymer molecules may freely enter and leave a formed capsid (left). The polymers inside the capsids are then in full equilibrium with the bulk polymer solution. In restricted equilibrium this is not the case and a fixed amount of polymer is trapped inside a capsid (right).

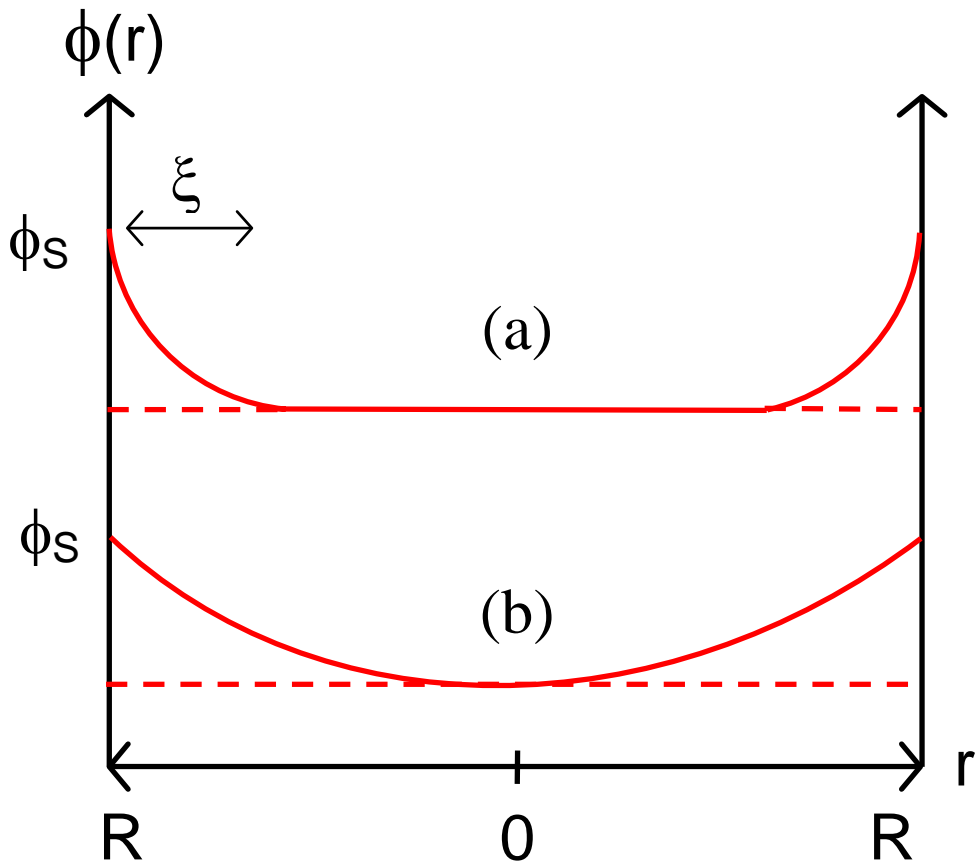

**Figure 2.** Polymer monomer concentration profiles $\phi(r)$ inside the capsid cavity as a function of the radial distance $r$. Fig.2a shows the exponential regime where the correlation length $\xi$ is less than the capsid radius R and Fig.2b shows the power-law regime where the correlation length exceeds R. In both regimes, most monomers are confined to a surface region with a thickness of order the extrapolation length d and a concentration of order $\phi_S$.

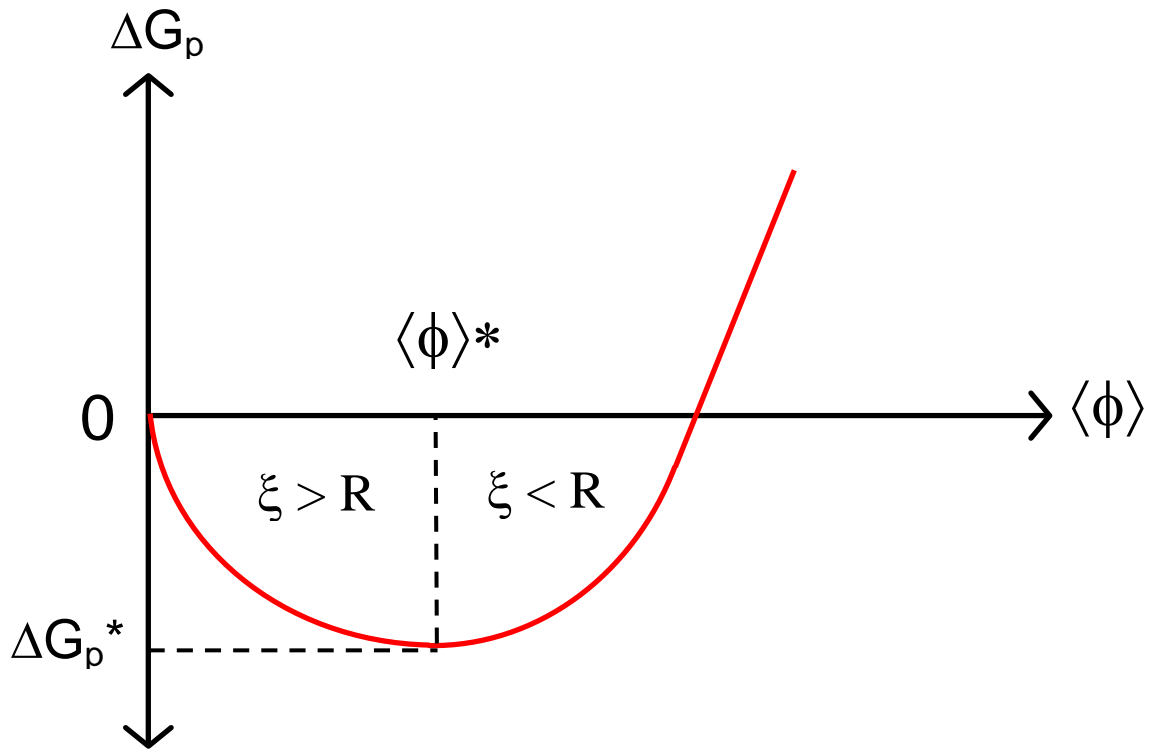

**Figure 3.** Co-assembly free energy $\Delta G_p$ of a capsid as a function of the mean polyelectrolyte monomer concentration $\langle \phi \rangle$ inside the capsid. Capsid/polyelectrolyte co-assembly requires $\Delta G_p$ to be negative. At the free energy minimum $\Delta G_p^*$, where filling concentration equals $\langle \phi \rangle^*$, the correlation length $\xi$ is comparable to the capsid radius $R$.

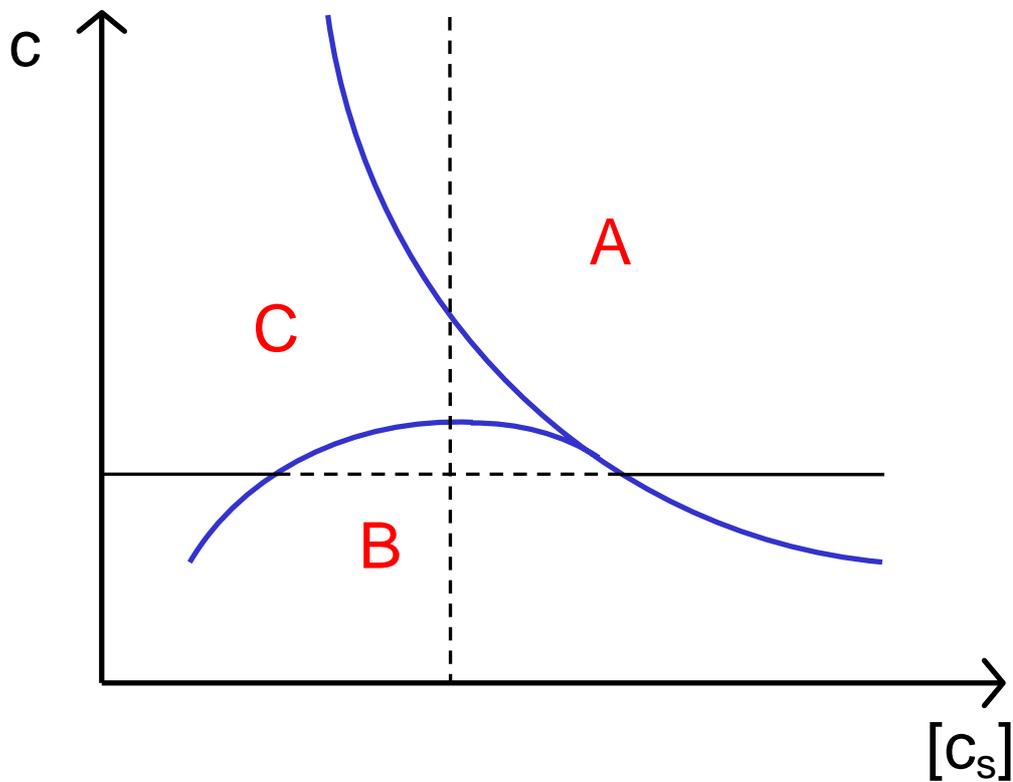

**Figure 4.** Self-assembly diagram of capsid proteins in the presence and absence of oppositely charged polyelectrolyte as a function of the salt concentration [$c_s$]. The full curves mark the critical subunit concentrations (CSC) beyond which capsid assembly takes place. In region A, empty capsids form in the absence of polyelectrolyte material. In region B, capsid-polyelectrolyte co-assembly takes place, while in region C neither empty nor filled capsids form. When we vary the salt concentration for fixed protein concentration – as indicated by the horizontal line – we encounter re-entrant phase behavior: capsids assemble for high, respectively, low salt concentrations producing empty, respectively, filled capsids, but over an intermediate interval of salt concentrations capsids are not stable. The intermediate interval terminates at the maximum of the CSC for filled capsids marking the *desorption transition* (vertical dashed line) beyond which polyelectrolyte molecules effectively lose their affinity for the capsid proteins.